\documentclass[letterpaper]{article} 
\usepackage{aaai2026} 
\usepackage{times}  
\usepackage{helvet}  
\usepackage{courier}  
\usepackage[hyphens]{url}  
\usepackage{graphicx} 
\usepackage{array}
\usepackage{booktabs}
\usepackage{natbib}  
\usepackage{caption} 
\usepackage{algorithm}
\usepackage{algorithmic}
 \usepackage{bm}
\usepackage{amsmath}
\usepackage{multirow} 
\usepackage{tabularx} 
\usepackage{newfloat}
\usepackage{listings}
\DeclareCaptionStyle{ruled}{labelfont=normalfont,labelsep=colon,strut=off} 
\floatstyle{ruled}
\newfloat{listing}{tb}{lst}{}
\floatname{listing}{Listing}
\title{Semantic Encryption: Secure and Effective Interaction with \\Cloud-based Large Language Models via Semantic Transformation
}
\author {
    Dong Chen\textsuperscript{\rm 1}\textsuperscript{\rm ,}\textsuperscript{\rm 2}\textsuperscript{\rm ,}\textsuperscript{\rm 3},
    Tong Yang\textsuperscript{\rm 1},
    Feipeng Zhai\textsuperscript{\rm 1},
    Pengpeng Ouyang\textsuperscript{\rm 1},
    Qidong Liu\textsuperscript{\rm 1}\textsuperscript{\rm ,}\textsuperscript{\rm 2}\textsuperscript{\rm ,}\textsuperscript{\rm 3},
    Yafei Li\textsuperscript{\rm 1}\textsuperscript{\rm ,}\textsuperscript{\rm 2}\textsuperscript{\rm ,}\textsuperscript{\rm 3},
    Chong Fu\textsuperscript{\rm 4},
    Mingliang Xu\textsuperscript{\rm 1}\textsuperscript{\rm ,}\textsuperscript{\rm 2}\textsuperscript{\rm ,}\textsuperscript{\rm 3}\textsuperscript{\rm ,}\thanks{Mingliang Xu is the corresponding author.},
}
\affiliations {
    \textsuperscript{\rm 1}The School of Computer and Artificial Intelligence of Zhengzhou University\\
    \textsuperscript{\rm 2}Engineering Research Center of Intelligent Swarm Systems, Ministry of Education\\
    \textsuperscript{\rm3}National Supercomputing Center In Zhengzhou\\
    \textsuperscript{\rm4}College of Computer Science and Technology of Zhejiang University\\
    chendongai@zzu.edu.cn, yangtong@gs.zzu.edu.cn, zhaifeipeng@stu.zzu.edu.cn, oy100253@gs.zzu.edu.cn, ieqdliu@zzu.edu.cn, ieyfli@zzu.edu.cn, fuchong@zju.edu.cn, iexumingliang@zzu.edu.cn
}

\usepackage{bibentry}

\begin{document}

\maketitle

\begin{abstract}
	The increasing adoption of Cloud-based Large Language Models (CLLMs) has raised significant concerns regarding data privacy during user interactions. While existing approaches primarily focus on encrypting sensitive information, they often overlook the logical structure of user inputs. This oversight can lead to reduced data utility and degraded performance of CLLMs. To address these limitations and enable secure yet effective interactions, we propose Semantic Encryption (SE)—a plug-and-play framework designed to preserve both privacy and utility. SE consists of two key components: Semantic Encoding and Semantic Decoding. In the encoding phase, a lightweight local model transforms the original user input into an alternative semantic context that maintains the original intent and logical structure while obfuscating sensitive information. This transformed input is then processed by the CLLM, which generates a response based on the transformed semantic context. To maintain a seamless user experience, the decoding phase will reconstruct the CLLM's response back into the original semantic context by referencing the locally stored user input. Extensive experimental evaluations demonstrate that SE effectively protects data privacy without compromising data utility or user experience, offering a practical solution for secure interaction with CLLMs. Particularly, the proposed SE demonstrates a significant improvement over the state-of-the-art InferDPT, surpassing it across various evaluated metrics and datasets.
\end{abstract}


\section{Introduction}

Cloud-based Large Language Models (CLLMs), which offer services such as data analysis through Application Programming Interfaces (APIs), are increasingly integrated into everyday life. However, transmitting data to the cloud via APIs for processing and analysis by CLLMs has raised significant concerns regarding data privacy. In particular, service providers may collect user data for model training purposes, further amplifying the risk of privacy leakages \cite{wang2024pandora,wu2024condefects,yang2023rethinking}.

\begin{figure}[t]
	\centering
	\includegraphics[scale=0.39]{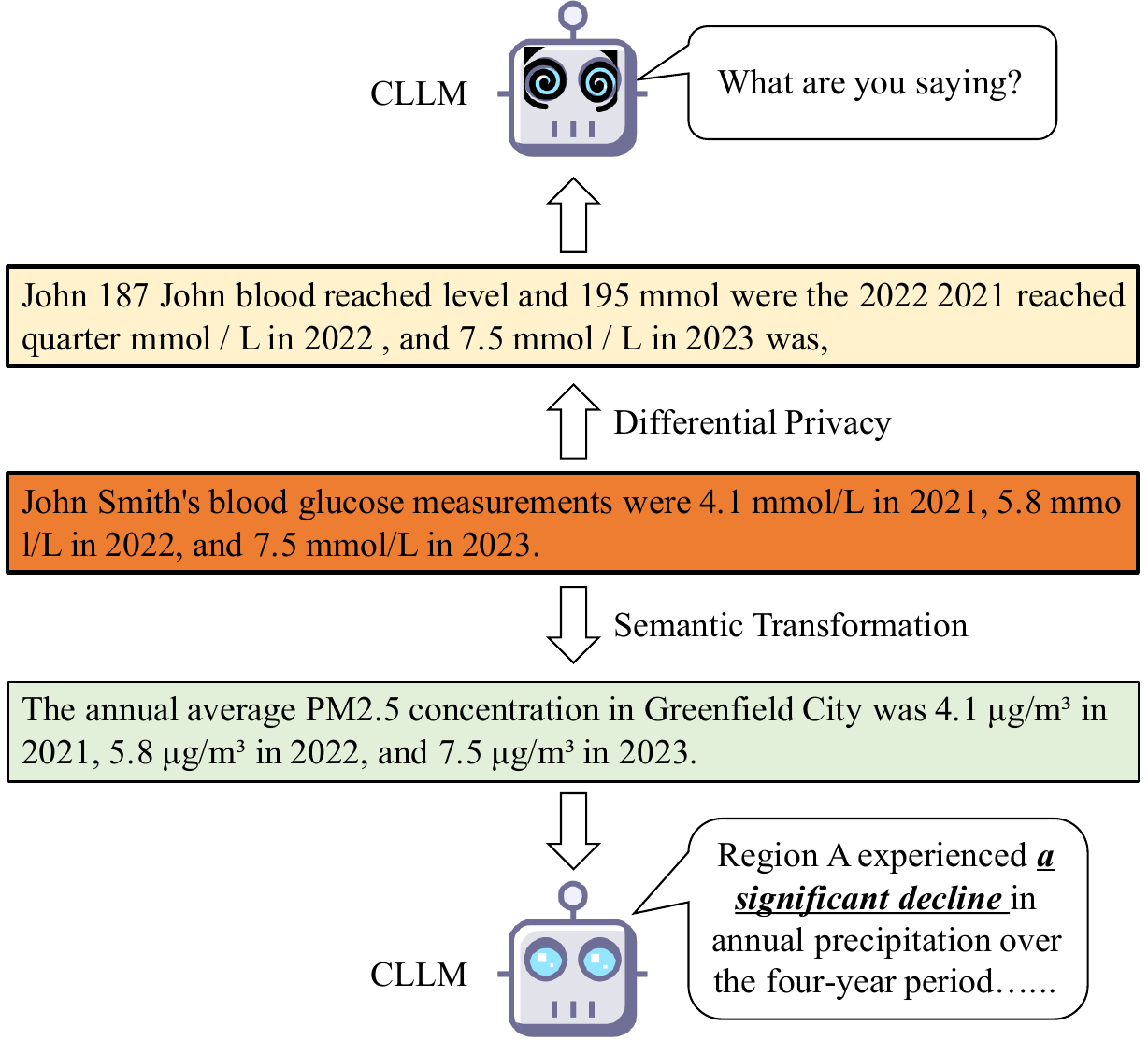}
	\caption{
A comparative case study of traditional encryption methods and the proposed semantic transformation in the context of interactions with CLLMs.	It is important to emphasize that the examples illustrated in the figure are entirely synthetic and do not contain any personal or sensitive information.
}
	\label{motivation}
\end{figure}
An increasing body of research has focused on protecting user data privacy during interactions with CLLMs, with particular emphasis on encryption-based techniques \cite{yan2024protecting, yao2024survey, feretzakis2024privacy}.
While encryption-based techniques such as differential privacy \cite{hoory2021learning, du2021dp} provide strong privacy guarantees by introducing randomness, they often come at the cost of reduced data utility, thereby impairing the CLLMs' ability to interpret and analyze user inputs effectively.
As illustrated in Figure \ref{motivation}, the orange box presents a patient's blood glucose test record. In the upper section of the figure, the input is encrypted with a differential privacy mechanism. 
Although the differential privacy method effectively safeguards data privacy, it substantially distorts the original intent and logical structure of the input, resulting in a complete loss of utility and hindering the CLLM’s ability to interpret user input.

Similar to the common practice of safeguarding privacy on public platforms by obscuring only personally identifiable information—such as blurring faces in photographs—it may be unnecessary to encrypt the entirety of user input to ensure data privacy \cite{chen2023fedaa,dong2024fadngs}, thereby helping to preserve the input’s logical structure and the user’s intent.
Motivated by this observation, we introduce semantic transformation that analyzes the original user input and transforms it into a logically consistent but semantically different representation. As illustrated in the light green box of Figure \ref{motivation}, semantic transformation converts the patient's blood glucose test record into a context representing annual average PM2.5 concentration record. 
This transformation can preserve the data utility for tasks such as trend analysis, while fully obscuring all the patient's information, thereby protecting data privacy.

Based on the idea of semantic transformation, we propose Semantic Encryption (SE) that comprises a Semantic Encoder and a Semantic Decoder. More specifically, the Semantic Encoder transforms the original input into an alternative semantic context while preserving its underlying logical structure. Correspondingly, the Semantic Decoder reconstructs the CLLMs' response based on the original input, restoring it to the original semantic context. To improve the deployability of the SE across heterogeneous devices, efficient and lightweight local models are utilized for both the encoder and decoder.
Furthermore, we propose Semantic Distillation, a technique that enables local models to effectively learn and replicate the semantic encoding and decoding capabilities of CLLMs. 
Since all user interactions with the CLLMs are preserved within their original semantic context, the operation of the SE remains virtually imperceptible to users.

\begin{figure*}[t]
	\centering
	\includegraphics[scale=0.65]{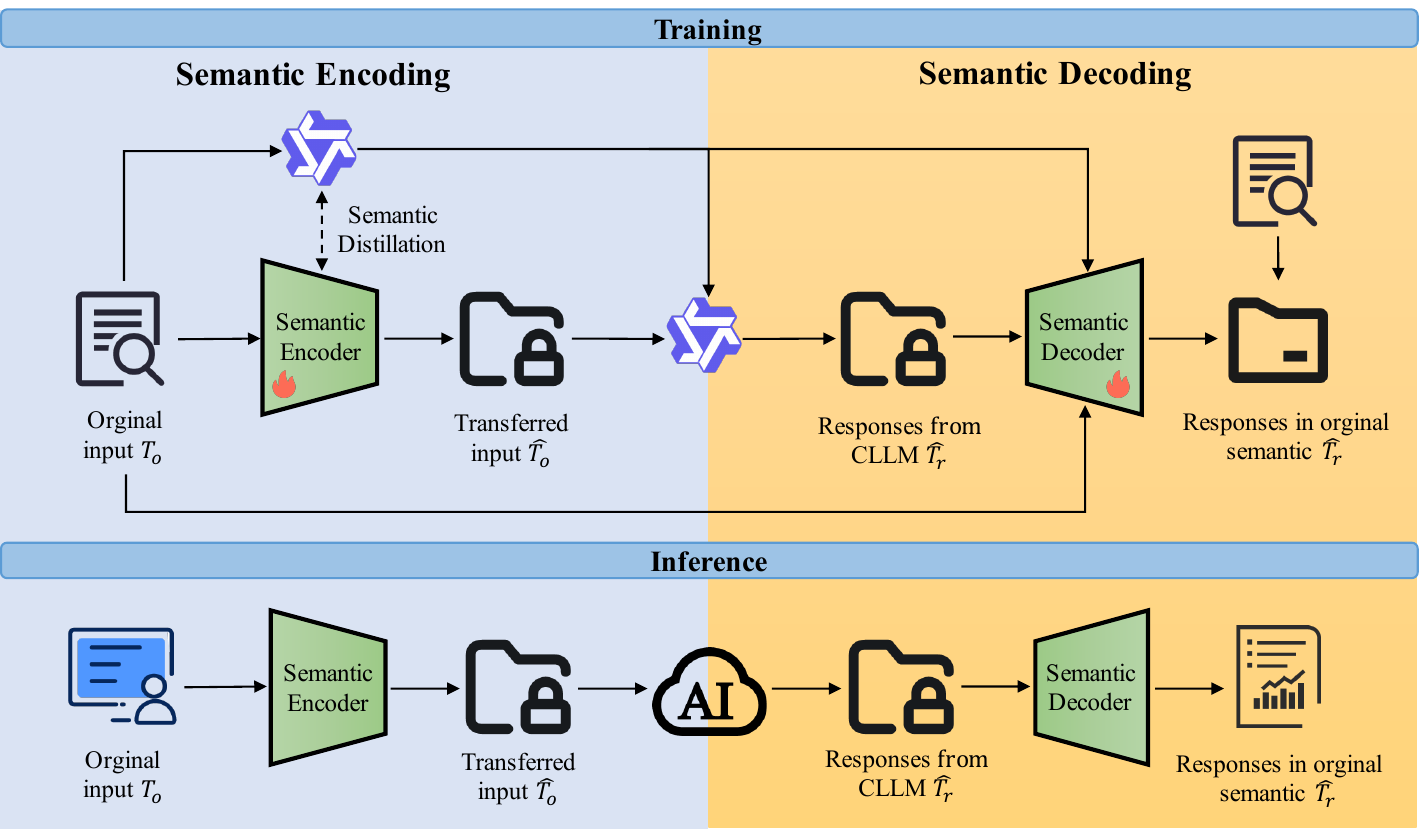}
	\caption{The proposed Semantic Encryption (SE) framework. SE consists of Semantic Encoding and Semantic Decoding, which focus on performance and user experience, respectively.}
	\label{pipline}
\end{figure*}

This paper proposes a method for protecting data privacy while simultaneously preserving data utility and ensuring a seamless user experience during interactions with CLLMs. The primary contributions of this work are summarized as follows:
\begin{itemize}
    \item We discuss the challenges of balancing data privacy and data utility in interactions with CLLMs, which traditional encryption methods fail to overcome.
    \item We propose SE that protects data privacy while preserving data utility by transforming the user's input into semantically distinct yet logically equivalent contexts.
    \item Extensive experiments demonstrate that the proposed method effectively protects data privacy, preserves data utility, and maintains user experience. 
\end{itemize}

\section{Related Work}

How to protect user privacy during interactions with CLLMs is gradually becoming a hot topic in current research. 
PrivacyRestore trains restoration vectors for each privacy span to alleviate insufficient privacy protection with performance degradation \cite{zeng2024privacyrestore}. 
Some studies protect data privacy by modifying user input keywords using local differential privacy mechanisms; however, this often leads to a degradation in data utility \cite{li2023privacy,hoory2021learning,du2021dp}. 
InferDPT \cite{tong2025inferdpt} leverages differential privacy to safeguard data privacy while concurrently training a decoder to reconstruct the encrypted content. Although this approach achieves a degree of balance between data utility and privacy preservation, it inevitably incurs degradation in the logical structure of inputs.

Pretrained large models possess extensive prior knowledge and strong reasoning capabilities but face challenges in being deployed across a wide range of devices \cite{chen2024data,chen2024improving}. In contrast, although small models have limited performance, they are easier to deploy. Thus, some studies have attempted to develop various techniques to enable small models to acquire the capabilities of pretrained large models \cite{fang2025knowledge,chen2025kka}.

In this paper, we propose a plug-and-play framework where local small models transform the original input from users and reponses from CLLMs.

\section{Methodology}
As illustrated in Figure \ref{pipline}, the proposed Semantic Encryption (SE) framework enhances data privacy by transforming user inputs into a logically consistent but semantically different representation, thereby preserving both privacy and utility. The response from the CLLM is subsequently mapped back to the original semantic context with a Semantic Decoder, enabling seamless and effective user interaction. In order to support a smooth user experience, Semantic Distillation is introduced within the SE, allowing a lightweight local model to approximate the semantic transformation capabilities.

\subsection{Semantic Encoding}
The Semantic Encoder $F_{SE}$ is implemented with a lightweight local model to transform the original input $T_o$ into a logically consistent but semantically different representation $\widehat{T_o}$ in other contexts. 
However, effectively mapping original semantic contexts to alternative ones necessitates extensive prior knowledge and strong logical reasoning capabilities, which lightweight models often lack. To address this limitation, we propose Semantic Distillation that extracts the prior knowledge and semantic transformation capabilities of CLLMs and distills them into a lightweight model.

Specifically, to enable $F_{SE}$ to acquire the semantic transformation capability of CLLMs $F_{CLLM}$ across diverse contexts, we first employ a random number generator to produce random number list $A$ with random lengths and values:
\begin{equation}
	\begin{aligned}
	&\mathbf{A} = [a_1, a_2, \dots, a_n], \\&\quad \text{where } n \sim \mathcal{U}(n_{\min}, n_{\max}), 
	\\&\quad a_i \sim \mathcal{U}(v_{\min}, v_{\max}), \quad i = 1, 2, \dots, n\\
	\end{aligned}
	\label{Eq: Rnum}
	\end{equation}
where the list length $n$ follows a discrete uniform distribution $\mathcal{U}$ within the range $[n_{\min}, n_{\max}]$, each element $x_i$ is randomly sampled from a given value range $[v_{\min}, v_{\max}]$, following a uniform distribution. 
The generated random number list, composed of various numerical combinations, is designed to stimulate the CLLM to produce a wide range of semantic contexts, thereby exploring more possibilities for semantic transformation.

Then, $F_{CLLM}$ can generate original input $T_o$ based on the list $A$:
\begin{equation}
T_o=F_{CLLM}(A,P_0)
\label{Eq: To}
\end{equation}
where $P_0$ is the prompt for generating $T_o$.
Intuitively, the greater the diversity of $T_o$, the more beneficial it is for training a general-purpose semantic encoder. However, due to limited computational resources, this paper focuses solely on task-specific semantic encryption. That is, if existing datasets for different tasks are available, $A$ in Equation \ref{Eq: To} will no longer be necessary, as $T_o$ in this case directly serves as the training data.

Based on the CLLM's extensive prior knowledge and powerful reasoning capabilities, the $T_o$ can be transformed to alternative semantic contexts.
\begin{equation}
\widehat{T_o}=F_{CLLM}(T_o)
\label{Eq: TR}
\end{equation}
By constructing $(T_o,\widehat{T_o})$ pairs with Equation \ref{Eq: TR}, we can extract the contexts knowledge and semantic transformation capability embedded in the CLLM. Subsequently, we fine-tune the lightweight local model with $(T_o,\widehat{T_o})$ pairs to enable effective transformation of user inputs by the following equation:
\begin{equation}
\min_{F_{SE}} \; \mathcal{L}(F_{SE}(T_o), \widehat{T_o})
\label{Eq: SE}
\end{equation}

After obtaining the Semantic Encoder by Equation \ref{Eq: SE}, we can send the transformed input to the CLLM to obtain corresponding response $\widehat{T_r}$:
\begin{equation}
\widehat{T_r}=F_{CLLM}(\widehat{T_o})
\label{Eq: T_r}
\end{equation}
 The response $\widehat{T_r}$ from the CLLM remains within the transformed semantic context.

\subsection{Semantic Decoding}
Semantic Encoding can protect data privacy and preserves data utility by transforming the original input $T_o$ to alternative semantic contexts $\widehat{T_o}$. However, this also causes the CLLM’s responses $\widehat{T_r}$ to remain within the transformed semantic context, potentially impacting the user experience. To address this issue, we introduce Semantic Decoding into the SE framework. By training an efficient Semantic Decoder, SE enables the transformation of the CLLM's responses $\widehat{T_r}$ back to the original semantic context $T_r$. 

Specifically, similar to Semantic Encoding, we provide the CLLM $F_{CLLM}$ with the original input $T_o$ and response $\widehat{T_r}$, enabling it to restore the response $\widehat{T_r}$ to the original semantic context. This process can be formally expressed as: 
\begin{equation}
T_r=F_{CLLM}(T_o,\widehat{T_r})
\label{Eq: DE_r}
\end{equation}
Subsequently, based on the $(T_o,\widehat{T_o},\widehat{T_r},T_r)$ quadruple, we train the Semantic Decoder $F_{SD}$ by: 
\begin{equation}
\min_{F_{SD}} \; \mathcal{L}(F_{SD}(T_o,\widehat{T_o},\widehat{T_r}), T_r)
\label{Eq: SD}
\end{equation}
Similar to $T_o$ in Equation \ref{Eq: To}, if datasets for different tasks are available, $T_r$ in Equation \ref{Eq: SD} can be directly replaced by the labels provided in the datasets.

The whole training process of SE is summarized in Algorithm \ref{Al: SE}.
\begin{algorithm}
	\caption{Training phase of Semantic Encryption.}
	\begin{algorithmic}
		\STATE Generate diverse context representations $\widehat{T_o}$ with CLLMs based on the training data $T_o$.
		\STATE Train Semantic Encoder by Equation \ref{Eq: SE} with $T_o$ and $\widehat{T_o}$.
		\STATE Process and analyze $\widehat{T_o}$ with the CLLM to obtain the response $\widehat{T_r}$.
		\STATE Treat labels $T_r$ in the training data as response in the original semantic context.
		\STATE Train Semantic Decoder by Equation \ref{Eq: SD} with $T_o$, $\widehat{T_o}$,  $\widehat{T_r}$ and $T_r$.
	\end{algorithmic}
	\label{Al: SE}
\end{algorithm}

\subsection{Can SE protect data privacy?}
In this section, we demonstrate that semantic transformation achieved through SE can effectively protect data privacy in the sense of Shannon.

Let $M$ denote the orginal context, i.e. the set of all possible representations in context $H$. Any $m \in M$ represents one such description. $N$ is the numerical data (e.g. measurements or statistics) that remain unchanged during transformation.
 $C$ denotes the ciphertext space, i.e. the set of all possible representations in context $B$.
We define a key space $K$, where each key $k \in K$ specifies a bijective ``semantic mapping'', i.e. $|K|=|C|$. Besides, we interpret $\Phi_k$ as the encryption function under key $k$, and denote:
\begin{equation}
E_k: M \rightarrow C
\end{equation}
where $k$ is chosen uniformly at random from $\mathcal{K}$.

For any fixed $m \in M$ and  $c \in C$,
\begin{equation}
\begin{aligned}
\Pr\bigl(M=m\mid C=c\bigr)&= \frac{\Pr(E_k(m)=c)\Pr(M=m)}{\Pr(C=c)}\\&=\Pr(M=m)
\end{aligned}
\end{equation}
so the ciphertext $C$ reveals no information about $M$. Furthermore, since $N$ is independent of the mapping $K$ and is revealed unchanged in this paper, the joint output $(C,N)$ satisfies
\begin{equation}
I\bigl(M;\,C,N\bigr) = 0
\label{Eq: I}
\end{equation}
where $I(\cdot;\cdot)$ denotes mutual information. Thus, according to \ref{Eq: I}, the proposed SE can protect data privacy for the original content.

\section{Experiments}
In our experiments, we aim to
(1) validate that SE can preserve data utility,
(2) validate that SE can maintain the user experience, 
(3) validate that SE can protect data privacy,
(4) demonstrate the specific workflow of SE through a case study,

\subsection{Datasets and Settings}

In this work, we conduct experiments on three mathematical reasoning datasets and one natural language inference (NLI) dataset:

\noindent  Gsm8K \cite{cobbe2021gsm8k}: Gsm8K is a high-quality benchmark dataset comprising elementary school-level math word problems, with 7,473 training samples and 1,319 testing samples. Each problem necessitates multi-step arithmetic reasoning, typically involving between 2 and 8 steps.

\noindent OrcaMath \cite{mitra2024orcamath}: OrcaMath is a synthetic dataset comprising mathematics problems reformulated using the GPT-4-Turbo model within the Agent-Instruct framework. From this dataset, 10,000 samples are selected for training and 5,000 samples for testing.

\noindent MetaMath \cite{yu2023metamath}: MetaMath is an augmented dataset derived from Gsm8K and MATH \cite{hendrycks2021measuring} through techniques such as problem restatement and reverse reasoning, comprising a total of 395,000 samples. From this dataset, 10,000 samples are used for training and 5,000 for testing.

\noindent ANLI \cite{nie2020adversarial}: ANLI is a NLI dataset constructed with multi-round human-and-model-in-the-loop adversarial training. It comprises three label categories: entailment, neutral, and contradiction. From this dataset, 15,000 samples are used for training, while 3,200 samples are used for testing.

We compare the proposed Semantic Encryption (SE) method with several differential privacy-based methods, including SANTEXT, SANTEXT$+$ \cite{yue2021differential}, CUSTEXT, and CUSTEXT$+$ \cite{chen2023customized}, as well as HaS \cite{chen2023hide} and InferDPT \cite{tong2025inferdpt}, which employs an encoder-decoder architecture.
In all the following experiments, the CLLM is Qwen-Plus, and both the Semantic Encoder and Semantic Decoder in SE are Qwen3-0.6B \cite{qwen3}. We train models with LoRA \cite{hu2022lora}, the rank is 8, the learning rate is 2e-5, the batch size is 2. All experiments are conducted on a single NVIDIA RTX A6000 GPU. For more details on the experimental settings, please refer to the accompanying code. All prompts used in the experiments are provided in the Appendix A.
The code and data for the proposed method are provided for research purpose \footnote{Code is included in the supplemental material. Code will be released upon the paper acceptance.}.

\begin{table}[t]
	\centering
	\caption{Data utility after user input encrypted by different privacy protection methods. The values in the table represent the accuracy of the CLLM in handling encrypted queries. For Gsm8K and Metamath, the values denote the probability that the CLLM generates the correct output. For ANLI, the values indicate the probability that the CLLM correctly identifies the underlying logical relationship. 
	}
	\label{Table: data utility}
	\begin{tabular}{lccc}
		\toprule
		Method & Gsm8K & Metamath & ANLI \\
		\midrule
		SANTEXT    & 0.00\% & 0.08\% & 33.38\% \\
		SANTEXT+   & 1.29\% & 3.36\% & 33.47\% \\
		CUSTEXT    & 4.09\% & 17.78\% & 34.59\% \\
		CUSTEXT+   & 7.66\% & 23.64\% & 42.53\% \\
		HaS        & 1.97\% & 16.46\% & 45.59\% \\
		SE(Ours)    & $\bm{83.02\%}$ & $\bm{84.08\%}$ & $\bm{55.78\%}$ \\
		\bottomrule
	\end{tabular}
\end{table}

\begin{table*}[t]
	\centering
	\caption{
		Impact of various privacy protection methods on user experience. All methods encrypt the user queries prior to processing by the same CLLM. The responses returned by the CLLM are subsequently post-processed according to the specifications of each method. Finally, the differences between the user-received responses and manual annotations are analyzed to evaluate the impact of the various encryption methods on user experience.
		BLEU denotes the average score from BLEU-1 to BLEU-4. BERTScore \cite{zhang2019bertscore} quantifies the overall similarity between the responses and manual annotations. 
	}
	\begin{tabularx}{\textwidth}{l|l|XXXXXX}
		\toprule
		Dataset & Method & BLEU$\uparrow$& METEOR$\uparrow$ & R-1$\uparrow$& R-2$\uparrow$& R-L$\uparrow$& BERTScore$\uparrow$ \\
		\midrule
		\multirow{7}{*}{Gsm8K} 
		& SANTEXT   & 0.0572 & 0.0509 & 0.0939 & 0.0012 & 0.0646 & 0.3676 \\
		& SANTEXT+  & 0.0728 & 0.0987 & 0.1323 & 0.0141 & 0.0899 & 0.4148 \\
		& CUSTEXT   & 0.0972 & 0.2061 & 0.2197 & 0.0486 & 0.1372 & 0.5166 \\
		& CUSTEXT+  & 0.1146 & 0.2793 & 0.2869 & 0.0799 & 0.1741 & 0.5758 \\
		& HaS       & 0.1299 & 0.2716 & 0.3713 & 0.1120 & 0.2385 & 0.6293 \\
		& InferDPT  & 0.1947 & 0.3687 & 0.4996 & 0.2139& 0.3443 & 0.7149 \\
		& SE (ours) & $\bm{0.2294}$ & $\bm{0.4368}$ & $\bm{0.5380}$ & $\bm{0.2524}$ & $\bm{0.3682}$ & $\bm{0.7243}$ \\
		\midrule
		\multirow{7}{*}{OrcaMath} 
		& SANTEXT   & 0.0133 & 0.0379 & 0.1255 & 0.0023 & 0.0609 & 0.3827 \\
		& SANTEXT+  & 0.0452 & 0.0898 & 0.1859 & 0.0282 & 0.1019 & 0.4390 \\
		& CUSTEXT   & 0.1216 & 0.2266 & 0.3340 & 0.1032 & 0.1936 & 0.5757 \\
		& CUSTEXT+  & 0.1687 & 0.3091 & 0.4173 & 0.1590 & 0.2449 & 0.6354 \\
		& HaS       & 0.2019 & 0.3171 & 0.4933 & 0.2077 & 0.2850 & 0.6628 \\
		& InferDPT  & 0.3187 & 0.4443 & 0.6321 & 0.3628 & 0.4129& 0.7516 \\
		& SE (ours) & $\bm{0.3638}$ & $\bm{0.4839}$ & $\bm{0.6771}$ & $\bm{0.4161}$ & $\bm{0.4614}$ & $\bm{0.7719}$ \\
		\midrule
		\multirow{7}{*}{Metamath} 
		& SANTEXT   & 0.0276 & 0.0460 & 0.1274 & 0.0026 & 0.0679 & 0.3732 \\
		& SANTEXT+  & 0.0712 & 0.1248 & 0.2147 & 0.0394 & 0.1184 & 0.4537 \\
		& CUSTEXT   & 0.1304 & 0.2347 & 0.3112 & 0.1005 & 0.1882 & 0.5622 \\
		& CUSTEXT+  & 0.1691 & 0.3184 & 0.3939 & 0.1523 & 0.2413 & 0.6271 \\
		& HaS       & 0.2288 & 0.3785 & 0.5176 & 0.2547 & 0.3335 & 0.6859 \\
		& InferDPT  & 0.3666 & 0.5056 & 0.6528 & 0.4202& 0.4815& 0.7754 \\
		& SE (ours) & $\bm{0.4272}$ & $\bm{0.5594}$ & $\bm{0.6942}$ & $\bm{0.4767}$ & $\bm{0.5268}$ & $\bm{0.7964}$ \\
		\bottomrule
	\end{tabularx}
	\label{Table: user experience}
\end{table*}

\subsection{Data Utility of Encrypted User Input for CLLMs}
In this section, we evaluate the effectiveness of CLLM in handling queries encrypted by different methods across various tasks to verify that SE can preserve data utility.

As shown in Table \ref{Table: data utility}, SE consistently achieves the best performance across both mathematical reasoning and NLI datasets. In particular, on the two mathematical reasoning benchmarks, SE surpasses the second-best baseline, CUSTEXT+, by 75.36\% on Gsm8K and 60.44\% on MetaMath. This significant improvement can be attributed to SE’s ability to transform the original input into semantically distinct yet logically equivalent contexts. In other words, SE effectively preserves critical components of mathematical reasoning, enabling the CLLM to accurately perform tasks based on encrypted inputs. In contrast, other methods employ random or rule-based substitutions of the user’s original input, which undermines the preservation of the user’s intention. 
On the other hand, although SE exhibits less pronounced advantages on the ANLI task compared to mathematical reasoning tasks, it still demonstrates its effectiveness on relatively simpler tasks. These results indicate that SE is more effective in contexts that rely on the reasoning capabilities.

\begin{table*}[htbp]
	\centering
	\caption{Quantitative comparison of privacy protection results across different methods.}
	\begin{tabularx}{\textwidth}{XXXXXX}
		\toprule
		\textbf{Dataset} & \textbf{Method} & \textbf{BLEU$\downarrow$} & \textbf{METEOR$\downarrow$} & \textbf{R-1$\downarrow$} & \textbf{R-2$\downarrow$}  \\
		\midrule
		\multirow{6}{*}{Gsm8K} 
		& SANTEXT     & 0.0492 & 0.0235 & 0.0228 & 0.0001  \\
		& SANTEXT+    & 0.2995 & 0.4919 & 0.5329 & 0.2958   \\
		& CUSTEXT     & 0.2153 & 0.3852 & 0.3782 & 0.2073   \\
		& CUSTEXT+    & 0.4347 & 0.6788 & 0.6273 & 0.4359   \\
		& HaS         & 0.6134 & 0.7941 & 0.8231 & 0.6664   \\
		& SE (Ours)    & 0.6109 & 0.7597 & 0.7954 & 0.6027   \\
		\midrule
		\multirow{6}{*}{OrcaMath} 
		& SANTEXT     & 0.0529 & 0.0238 & 0.0251 & 0.0001  \\
		& SANTEXT+    & 0.3035 & 0.4841 & 0.5258 & 0.2990   \\
		& CUSTEXT     & 0.2170 & 0.3863 & 0.3761 & 0.2100   \\
		& CUSTEXT+    & 0.4383 & 0.6808 & 0.6216 & 0.4380  \\
		& HaS         & 0.6301 & 0.8043 & 0.8307 & 0.6830  \\
		& SE (Ours)    & 0.5951 & 0.7517 & 0.7918 & 0.5909   \\
		\midrule
		\multirow{6}{*}{MetaMath} 
		& SANTEXT     & 0.0407 & 0.0207 & 0.0235 & 0.0014  \\
		& SANTEXT+    & 0.3287 & 0.5108 & 0.5614 & 0.3358   \\
		& CUSTEXT     & 0.3023 & 0.4514 & 0.4280 & 0.2744  \\
		& CUSTEXT+    & 0.5254 & 0.7325 & 0.6716 & 0.5102  \\
		& HaS         & 0.6955 & 0.8237 & 0.8481 & 0.7212 \\
		& SE (Ours)    & 0.7169 & 0.8167 & 0.8411 & 0.6929   \\
		\bottomrule
	\end{tabularx}
	\label{Table: Data Privacy}
\end{table*}

\subsection{User Experience with CLLM's Responses}

In the Table \ref{Table: user experience}, the encrypted inputs are initially processed by the CLLM. Subsequently, each method applies an additional post-processing step to the CLLM’s output to generate the final responses presented to the user.
By comparing the final responses with corresponding manual annotations, we can evaluate the impact of each method on user experience.
It can be observed that methods incorporating a decoding process, such as HaS, InferDPT and the proposed SE, consistently outperform differential privacy-based methods, including SANTEXT, SANTEXT+, CUSTEXT, and CUSTEXT+.
This is primarily because differential privacy introduces random noise to achieve encryption, which can cause a substantial semantic divergence between the encrypted and original inputs, thereby leading to a marked deviation between the CLLM's output and the intended result.
In contrast, HaS, InferDPT and SE establish a mapping between the original and encrypted content, enabling the transformation of the CLLM’s output back into a context that is more readily interpretable by users.
InferDPT leverages a designed differential privacy strategy to encrypt the user's original input and subsequently trains a decoder to reconstruct it. However, InferDPT's encryption process inherently leads to the irreversible loss of certain useful information.
Similarly, HaS relies exclusively on the substitution of sensitive terms, which often disregards the user's original intent and the logical structure of the input, potentially impairing the performance of the CLLM and adversely affecting the user experience.
In comparison, SE transforms user input into a semantically similar context that preserves logical coherence, and leverages a semantic decoding module to reconstruct the CLLM’s response. As a result, SE consistently outperforms all baselines across all datasets and evaluation metrics, which demonstrates that SE can effectively maintain user experience.

It is important to note that our implementation of InferDPT adopts CUSTEXT+ for the encryption process. Consequently, its performance with respect to data utility and data privacy is consistent with that of CUSTEXT+. Therefore, InferDPT is only included in the Table \ref{Table: user experience}.

\begin{figure}[t]
	\centering
	\includegraphics[scale=0.3]{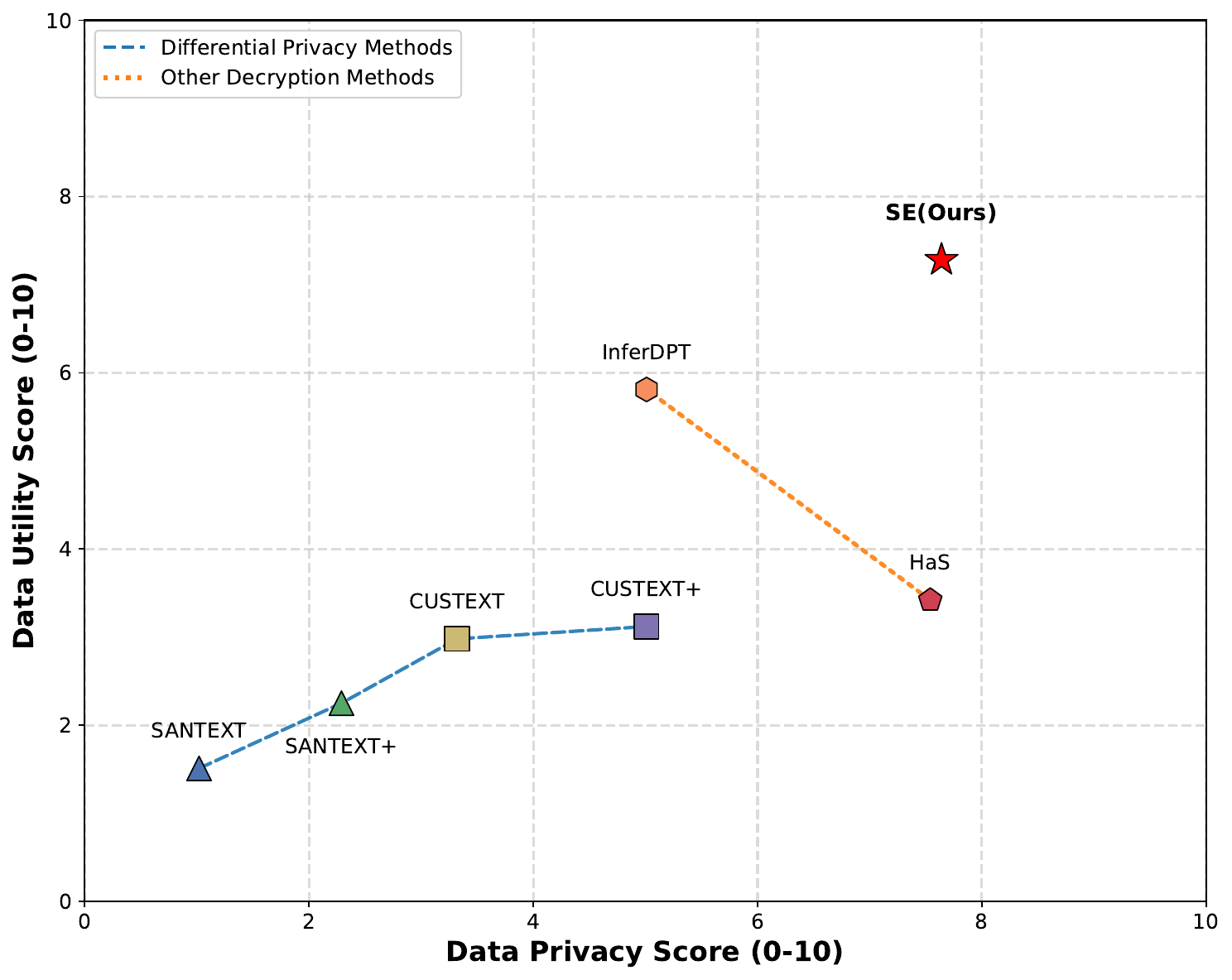}
	\caption{The average evaluation results of the LLM applied to various encryption methods across three mathematical reasoning datasets.
	}
	\label{LLM evaluatio}
\end{figure}

\begin{figure*}[t]
	\centering
	\includegraphics[scale=0.54]{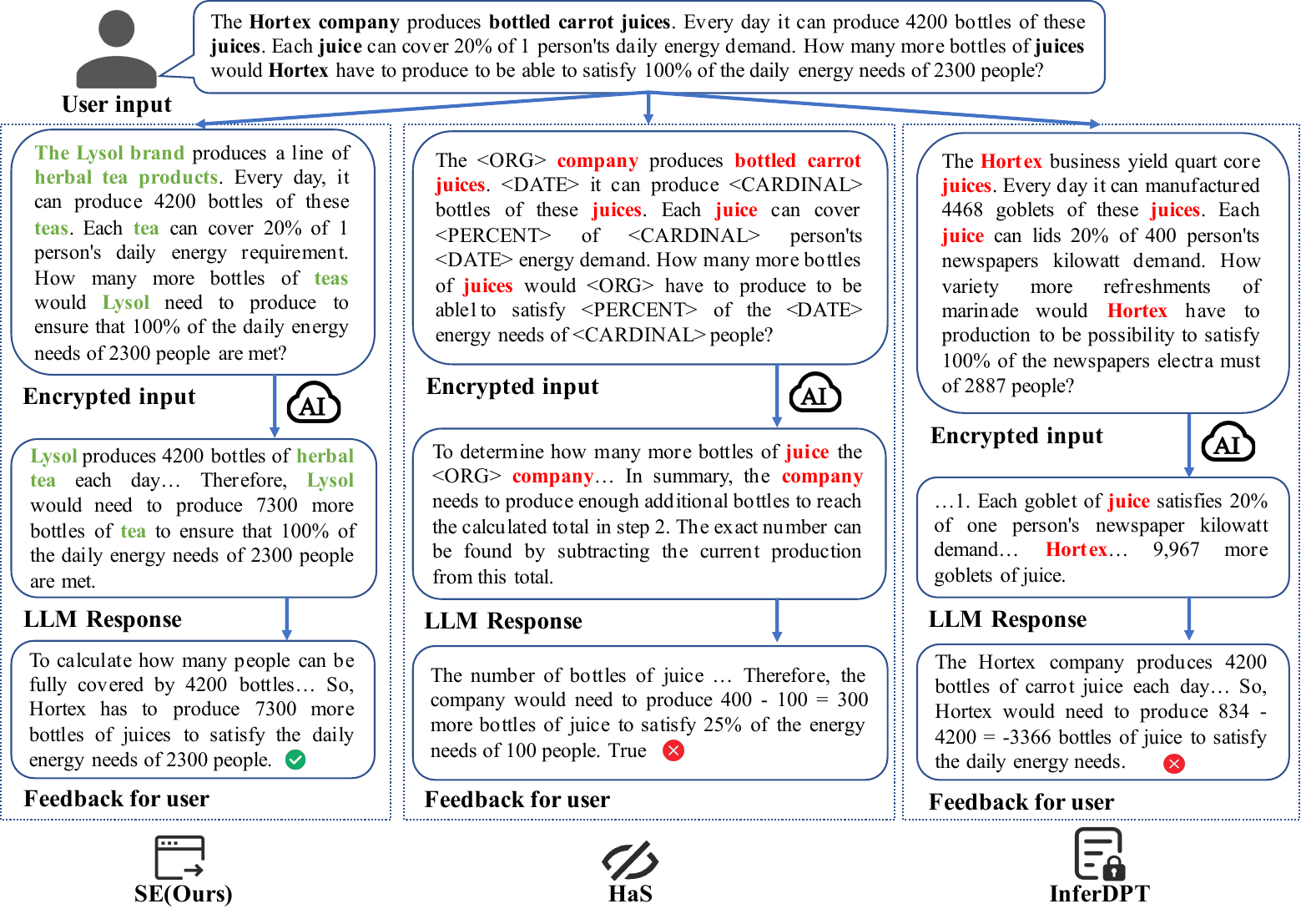}
	\caption{The case study for SE and Has. Bold black font denotes privacy-sensitive information requiring protection, red indicates instances of privacy leakage, and green signifies effective privacy protection.}
	\label{case study}
\end{figure*}

\subsection{Data Privacy of Encrypted User Input}
In this section, we follow the prior work \cite{li2025priv,chen2023hide} to evaluate the privacy protection capabilities of different methods by assessing the similarity between the user’s original input and its encrypted counterpart. The similarity is quantified using standard metrics such as BLEU and ROUGE. Intuitively, lower BLEU and ROUGE scores indicate stronger privacy protection. Related results are reported in Table \ref{Table: Data Privacy}.

The results presented in Table \ref{Table: Data Privacy} indicate that traditional differential privacy-based methods, such as SANTEXT, SANTEXT+, CUSTEXT, and CUSTEXT+, are highly effective in preserving data privacy. This effectiveness is primarily attributed to the introduction of varying levels of noise into the original user inputs, whereby higher noise magnitudes lead to greater divergence between the encrypted content and the original data.
However, as demonstrated by the results in Tables \ref{Table: data utility} and \ref{Table: user experience}, the introduction of noise significantly compromises data utility, leading to notable degradation in both CLLM performance and user experience. In contrast, other decryption approaches such as HaS and the proposed SE, which adopt encoder–decoder architectures, offer a more favorable balance between privacy protection and data utility. This is achieved by selectively transforming only the sensitive components of user input, thereby avoiding unnecessary obfuscation of non-sensitive segments. Furthermore, compared to HaS—which operates by selectively masking identified sensitive terms—SE provides stronger privacy guarantees in most contexts. For instance, on the OrcaMath dataset under the R-2 metric, the similarity between SE-encrypted text and the original input is 0.0921 lower than that of HaS, indicating a greater degree of semantic transformation and, consequently, enhanced privacy protection. 

\subsection{Large Language Model Evaluation}
In this section, we evaluate various encryption methods using the advanced LLM Qwen-Plus. Detailed evaluation metrics and prompts are provided in the Appendix A. The corresponding results are presented in Figure \ref{LLM evaluatio}.
Compared to the results presented in Table \ref{Table: Data Privacy}, this experiment places greater emphasis on privacy protection from the user’s perspective, including metrics such as Logical Structure Preservation. By effectively replacing irrelevant information while maintaining the original input’s logical structure, the proposed SE achieves superior performance in data privacy protection. Furthermore, evaluations of response quality from the user’s perspective such as Logical Reasonableness, further validate the suitability of the proposed method for interactive contexts involving CLLMs.

\subsection{Case Study}
We present the workflows of SE and the two best-performing baselines, HaS and InferDPT, in Figure \ref{case study}.
HaS protects private information through keyword substitution. However, its encrypted input reveals limitations in effectively securing all critical elements; for instance, terms like “juices” remain unencrypted, potentially disclosing that the input relates to a juice manufacturing company. 
As for InferDPT, it encrypts the user input with differential privacy. Nevertheless, its effectiveness in protecting data privacy is inconsistent—for instance, it fails to encrypt critical information such as company name in this case. Moreover, HaS and InferDPT compromises data utility, preventing the CLLM from responding based on the correct logical relationships.
In contrast, SE treats numerical values as shareable information and transforms critical elements into alternative semantic contexts, thereby facilitating the effective execution of relevant analytical tasks.
Besides, during the user feedback stage, SE restores the CLLM's response into the original semantic context and returns them to the user. Throughout the interaction with the CLLMs, users will remain unaware of the presence of the SE framework. 
 
The Appendix B further reports the response accuracy of the three methods. Notably, SE substantially outperforms the baselines, particularly on the GSM8K dataset, achieving accuracy gains of 34.34\% and 63.83\% over InferDPT and HaS, respectively.

\section{Conclusion}
In light of the growing importance of interactions with CLLMs, traditional encryption techniques such as differential privacy offer protection for sensitive data but usually at the cost of data utility. 
The degradation of data utility impairs the ability of CLLMs to generate satisfactory responses to user queries, as traditional encryption techniques hinder the CLLMs' understanding of encrypted user inputs.
 To overcome these limitations, we introduce Semantic Encryption, a novel framework consists of Semantic Encoding and Semantic Decoding. 
 For Semantic Encoding, a semantic encoder analyzes the user’s original input and transforms it into an alternative semantic context that preserves the original logical structure. 
 After the encrypted input is processed by the CLLM, a semantic decoder is employed to map the CLLM’s response back to the original semantic context. 
This end-to-end process ensures data privacy throughout the interaction, while preserving high response quality and delivering a seamless user experience.

\bibliography{aaai2026}

\end{document}